\documentclass[prd,nofootinbib,preprint,superscriptaddress]{revtex4}
\pdfoutput=1

\usepackage{amsmath,amssymb}
\usepackage{epsfig}
\usepackage{graphicx}
\usepackage[usenames,dvipsnames]{color}
\usepackage{subfigure}
\usepackage{slashed}
\usepackage[colorlinks,citecolor=blue]{hyperref}
\usepackage{color}
{}
{}

\begin{document}
	\title{Boosted Self-Interacting Dark Matter and XENON1T Excess}
	
	\author{Debasish Borah}
	\email{dborah@iitg.ac.in}
	\affiliation{Department of Physics, Indian Institute of Technology Guwahati, Assam 781039, India}
	
	\author{Manoranjan Dutta}
	\email{ph18resch11007@iith.ac.in}
	\affiliation{Department of Physics, Indian Institute of Technology Hyderabad, Kandi, Sangareddy 502285, Telangana, India}
	
	\author{Satyabrata Mahapatra}
	\email{ph18resch11001@iith.ac.in}
	\affiliation{Department of Physics, Indian Institute of Technology Hyderabad, Kandi, Sangareddy 502285, Telangana, India}

	\author{Narendra Sahu}
	\email{nsahu@phy.iith.ac.in}
	\affiliation{Department of Physics, Indian Institute of Technology Hyderabad, Kandi, Sangareddy 502285, Telangana, India}
	
	\begin{abstract}
		We present a self-interacting boosted dark matter (DM) scenario as a possible explanation of the recently reported excess of electron recoil events by the XENON1T experiment. The Standard Model (SM) has been extended with two vector-like fermion singlets charged under a dark $U(1)_D$ gauge symmetry to describe the dark sector. While the presence of light vector boson mediator leads to sufficient DM self-interactions to address the small scale issues of cold dark matter, the model with sub-GeV scale DM can explain the XENON1T excess via elastic scattering of boosted DM component with electrons at the detector. Strong annihilation of DM into the light mediator leads to a suppressed thermal relic. A hybrid setup of dark freeze-out and non-thermal contribution from the late decay of a scalar can lead to correct relic abundance. We fit our model with XENON1T data and also find the final parameter space consistent with self-interaction of DM, DM-electron scattering rate, as well as astrophysical and cosmological observations. A tiny parameter space consistent with all these constraints and requirements can be further scrutinized in near-future experiments.
	\end{abstract}	
	\maketitle
	\section{Introduction}
	\label{intro}
	The current Standard Model(SM) of particle physics is extremely successful in explaining the fundamental properties of elementary particles and their interactions in nature. However, it falls short in explaining some of the mysteries of the Universe. For instance, now there exist sufficient evidence from astrophysics and cosmology towards the presence of Dark Matter (DM) which is a non-luminous, non-baryonic form of matter that constitutes a significant portion of the whole universe \cite{ParticleDataGroup:2020ssz, Planck:2018vyg}. Precise measurement of anisotropies in the cosmic microwave background radiation (CMBR) by the Planck and WMAP like satellite-borne experiments predict the amount of DM in the present Universe to be around one-fourth ($26.8\%$) of the current energy density of the Universe.
	Conventionally the DM abundance is reported in terms of density parameter $\Omega_{\rm DM}$ and 
	$h = \text{Hubble Parameter}/(100 \;\text{km} ~\text{s}^{-1} 
	\text{Mpc}^{-1})$ as \cite{Planck:2018vyg}:
	$\Omega_{\text{DM}} h^2 = 0.120\pm 0.001$
	at 68\% CL. Data collected over a long period of time since the 1930s from observations of different galaxies and clusters also support this 
	number \cite{Zwicky:1933gu, Rubin:1970zza, Clowe:2006eq}.  Here it is worth mentioning that the estimate of the present DM abundance by Planck relies upon the 
	standard model of cosmology or ${\rm \Lambda CDM}$ model, where $\Lambda$ denotes the cosmological constant or dark energy and CDM refers to cold dark matter, a pressure-less or collision-less fluid, which is essential for structure formation. As the latter requires a gravitational potential well for ordinary matter to 
	collapse and form structures, CDM provides a seed for the creation of that potential well. Though $\Lambda$CDM model has been very successful in describing our 
	Universe at large scale $(\geq \mathcal{O}(\rm Mpc))$, at small scales, it faces challenges from observations like too-big-to-fail, missing satellite and 
	core-cusp problems. For recent reviews of such anomalies and possible solutions, refer to \cite{Tulin:2017ara, Bullock:2017xww}. One possible solution to 
	these puzzles can be self-interacting dark matter (SIDM) as an alternative to the collision-less CDM\footnote{See \cite{deLaix:1995vi} for earlier studies.} 
	which was first proposed by Spergel and Steinhardt \cite{Spergel:1999mh}.  The fascinating feature of SIDM is that it can solve the problems at small scales while being consistent with the observed CDM halos at large radii. The required self-interaction rate is often quantified in terms of the ratio of cross-section to 
	DM mass as $\sigma/m \sim 1 \; {\rm cm}^2/{\rm g} \approx 2 \times 10^{-24} \; {\rm cm}^2/{\rm GeV}$ \cite{Buckley:2009in, Feng:2009hw, Feng:2009mn, Loeb:2010gj, Zavala:2012us, Vogelsberger:2012ku}.

	Since we still don't have an answer to what DM actually is, as none of the SM particles has the properties that a DM particle is expected 
	to have, over the years people have resorted to several beyond standard model (BSM) scenarios. Among all such BSM frameworks, the weakly interacting massive 
	particle (WIMP) paradigm has been the most widely studied one where a DM candidate having interactions and mass in the typical electroweak regime naturally 
	satisfies the correct DM relic abundance through a thermal freeze-out mechanism- an astounding coincidence referred to as the \textit{WIMP Miracle}~\cite{Kolb:1990vq,Arcadi:2017kky}. However, such sizeable DM-SM interactions have not been observed yet at typical direct detection experiments like LUX, PandaX, XENON1T etc. which rely on DM-nucleon scattering events as a signal. Thus it has motivated the particle physics community to look for several viable alternatives to WIMP. Without giving up entirely on WIMP, one exciting possibility that has recently gained popularity is light DM around GeV or sub-GeV scale. While WIMP with electroweak type interactions has a lower bound on its mass, around a few GeV, known as the Lee-Weinberg bound \cite{Lee:1977ua}, one can relax these bound in specific models where additional light particles mediate DM-SM interactions.
	Also the required self-interaction cross-sections ($\sigma/m \sim 1 \; {\rm cm}^2/{\rm g} \approx 2 \times 10^{-24} \; {\rm cm}^2/{\rm GeV}$) can be naturally realized in models with light mediators. In such scenarios, self-interactions can be shown to be stronger for smaller DM velocities such that it can have a large impact on small scale structures while being consistent with usual CDM predictions at larger scales \cite{Buckley:2009in, Feng:2009hw, Feng:2009mn, Loeb:2010gj, Bringmann:2016din, Kaplinghat:2015aga, vandenAarssen:2012vpm, Tulin:2013teo}. From a particle physics point of view, such self-interactions can be naturally realized in Abelian gauge extensions of the SM where the new gauge boson is light. As the dark sector can not be completely hidden and there should be some coupling of the new mediator with SM particles as well to ensure that DM and SM sectors were in thermal equilibrium in the early Universe, the same coupling can also be probed at DM direct detection experiments \cite{Kaplinghat:2013yxa, DelNobile:2015uua}, and indeed one such possibility is the topic of this work.
	
	The sub-GeV scale DM with light mediators have recently received a lot of attention after XENON1T collaboration published their latest results in June 2020 where they have reported observation of an excess in electron recoil events over the background in the recoil energy $E_r$ in a range 1-7 keV, peaked around 2.4 keV\cite{XENON:2020rca}. While this excess can possibly be explained by solar axions at $3.5\sigma$ significance or neutrino magnetic moment at $3.2\sigma$ significance both these interpretations are in strong tension with stellar cooling constraints. While there is also room for possible tritium backgrounds in the detector, which XENON1T collaboration has neither confirmed nor ruled out so far, there have been several interesting new physics proposals in the literature. For example, see~ \cite{Takahashi:2020bpq,Alonso-Alvarez:2020cdv,Kannike:2020agf,Fornal:2020npv,Du:2020ybt,Ko:2020gdg, Su:2020zny,Harigaya:2020ckz, Borah:2020jzi,Choudhury:2020xui,Buttazzo:2020vfs, Bramante:2020zos, Bell:2020bes,  Borah:2020smw, Aboubrahim:2020iwb, Lee:2020wmh, Baek:2020owl, Shakeri_2020, Bally:2020yid, DelleRose:2020pbh, Ema:2020fit, Dutta:2021wbn} and references therein. The DM interpretations out of these examples, typically have a light mediator via which DM interacts with electrons. The recoil can occur either due to light boosted DM or inelastic up or down-scattering~\cite{Bell:2020bes, Lee:2020wmh, Baek:2020owl, Harigaya:2020ckz, Bramante:2020zos, Baryakhtar:2020rwy, Chao:2020yro, An:2020tcg, He:2020wjs, Choudhury:2020xui, Borah:2020jzi, Kim:2020aua, Shakeri_2020, Borah:2020smw, Keung:2020uew, Aboubrahim:2020iwb, He:2020sat, Choi:2020ysq, McKeen:2020vpf, Jho:2020sku, Alhazmi:2020fju, Jho:2020sku, Dutta:2021wbn, Das:2021lcr, Borah:2021jzu}. For further detection prospects of such boosted DM in different experiments, see~\cite{Alhazmi:2016qcs, Berger:2014sqa, Jho:2020sku, Kong:2014mia, Agashe:2014yua}.

	Thus, we realized that in a model with a light mediator, DM interpretation of XENON1T excess and self-interaction of DM can be simultaneously explained and this motivates us to propose a common platform to demonstrate that the self-interaction of DM arising via light mediators can also give rise the observed XENON1T excess. Hence this proposed framework provides a unique way of scrutinizing the SIDM parameter space at direct DM search experiments like XENON1T. There have been two such attempts so far trying to address XENON1T excess within the SIDM framework. In our earlier work \cite{Dutta:2021wbn}, we considered inelastic SIDM scattering off electrons, while in another recent work \cite{Baek:2021yos} considered the decay of an excited state into DM and a very light sub-eV vector mediator leading to a dark photo-electric effect. In the present work, we consider the possibility of boosted SIDM where heavier DM annihilates into the lighter one followed by the scattering of the latter off electrons at the XENON1T detector\footnote{Only boosted DM interpretation of XENON1T excess in the context of different models have been discussed in \cite{Kannike:2020agf,Fornal:2020npv,Du:2020ybt,Ko:2020gdg, McKeen:2020vpf, Jho:2020sku, Alhazmi:2020fju, Jho:2020sku, Das:2021lcr, Borah:2021jzu}. See \cite{Kim:2016zjx, Giudice:2017zke} for earlier works on this possibility.}. To be specific, in this scenario, the dark sector consists of two vector-like fermion singlets charged under an additional $U(1)_D$ gauge symmetry and the corresponding vector boson $Z'$ which mediates DM self-interactions is considered to be light to facilitate the required self-interactions at different scales. The mixing of this $Z'$ gauge boson with the $U(1)_Y$ gauge boson provides the necessary portal for DM direct detection, specifically the electron recoil events at XENON1T detector in this case. 
	
	In inelastic DM scenarios considered in~\cite{Dutta:2021wbn}, the off-diagonal vector coupling between the two DM candidates with the light gauge boson is the key feature that is essential to explain the $DM-e$ scattering via the down scattering of the heavier DM component. However, there is no off-diagonal coupling between the two DM candidates in the scenario considered in this paper. So unlike the earlier scenario, here, only elastic scattering of the DM with the electron is possible. Moreover, if one considers just the usual cold DM elastically scattering with the electron, then the recoil energy of the electron is of the order  $\mathcal{O}$(eV), which can not explain the keV range recoil excess observed at XENON1T. So to explain the XENON1T excess, it is required to consider a mechanism to impart a particular boost to a DM component which can elastically scatter off the electron giving a recoil of the order $\mathcal{O}$(keV). It is worth mentioning here that unlike inelastic DM scenarios where if the DM is in GeV scale, the local galactic DM density can provide enough DM flux to give rise to the reported electron recoil event rate at XENON1T detector, in boosted DM scenarios, obtaining a sufficient flux of the boosted DM component is a challenge. The need of an appropriate flux of the boosted DM that can explain the XENON1T electron recoil data forces one to consider lighter DM particles in the sub-GeV range. In addition, as we rely on the annihilation of the heavier DM component to the lighter one to generate the necessary boosted DM flux, this annihilation cross-section is also a crucial parameter in determining the flux. Also, to regenerate the reported electron recoil event rate, the boost or the velocity of the DM component is a decisive factor which is essentially determined by the mass difference between the two elastic SIDM components. To have a velocity of the boosted DM component in an approximate range $(0.05-0.1)c$ which can give a nice fit to the XENON1T data, the mass-splitting($\Delta M$) between the two DM components has to be such that $\Delta M/ M_{\rm BDM} \sim 10^{-3}$ and hence for a sub-GeV DM, the mass-splitting should be in the sub-MeV scale. This is also another significant difference from the inelastic DM scenarios where the mass-splitting has to be strictly around $2.4$ keV to explain the XENON1T excess.
	
	We analyze the DM parameter space consistent with velocity-dependent self-interaction rates that can explain the astrophysical data at the scale of clusters, galaxies and dwarf galaxies. Then confronting the SIDM parameter space with the observed XENON1T electron excess and other experimental and phenomenological bounds, we see that pure thermal relic of DM is insufficient to produce the observed relic, and therefore we consider a hybrid setup where both freeze-out and freeze-in mechanisms play crucial roles in generating DM relic. As discussed in the upcoming sections, invoking a singlet scalar that can decay into DM at late times helps generate the correct DM relic in such a hybrid setup.
	
	This paper is organized as follows. In section \ref{model}, we present our model followed by the analysis for dark matter self-interaction in section \ref{self_int}. In section \ref{dm_production}, we discuss the production mechanism of DM in the early Universe. The possible origin of XENON1T excess in our model via boosted DM scenario has been discussed in section \ref{xenon} and the detection prospects of the boosted DM at DM-nucleon scattering experiments are discussed in section~\ref{direct_search}. We finally summarize our results and conclude in section \ref{conclusion}.			
	
	\section{The Model}
	\label{model}
	\begin{table}[h!]
			\begin{tabular}{|c|c|c|c|}
				\hline \multicolumn{2}{|c}{Fields}&  \multicolumn{1}{|c|}{ $SU(3)_c \otimes SU(2)_L \otimes U(1)_Y \otimes  U(1)_D $  } \\ \hline
				{Fermion} 
				& $\chi_1$  & ~~1 ~~~~~~~~~~~1~~~~~~~~~~0~~~~~~~~~~ -1 \\
				[0.5em] 
				& $\chi_2$  & ~~1 ~~~~~~~~~~~1~~~~~~~~~~0~~~~~~~~~~ -1\\
				\hline
				{Scalars} & 
				$\Phi_1$ & ~~1 ~~~~~~~~~~~1~~~~~~~~~~0~~~~~~~~~~ 0 \\
				&	$\Phi_2$ & ~~1 ~~~~~~~~~~~1~~~~~~~~~~0~~~~~~~~~~ {0} \\
				\hline
			\end{tabular}
			\caption{BSM fields and their transformations under the gauge symmetry.}
			\label{tab:tab2}
		\end{table}	
		The matter particle content of the model apart from the SM ones are shown in table \ref{tab:tab2}. The Lagrangian with the interactions relevant for determining the DM abundance in the considered scenario is given by
		\begin{align}
			\label{Lagrangian}
			\mathcal{L}_{\rm DM} & \supset i ~\overline{\chi_i} \gamma^\mu~ D_\mu \chi_i - m_i \overline{\chi_i} \chi_i  - y_{i} \overline{\chi_i} \chi_i \Phi_1-y'_{i} ~\overline{\chi_i} \chi_i \Phi_2 + \frac{\epsilon}{2}B^{\alpha \beta}Y_{\alpha\beta}
		\end{align}
		where $D_\mu = \partial_\mu + i g' Z'_\mu$ and $B^{\alpha\beta}, Y_{\alpha \beta}$ are the field strength tensors of $U(1)_D, U(1)_Y$ respectively and $\epsilon$ is the kinetic mixing between them. The subscript $i=1,2$ corresponds to two different singlet fermions. We consider the mass and couplings of two singlet fermions in their diagonal mass basis. The singlet scalars $\Phi_1, \Phi_2$ are assumed not to acquire any vacuum expectation values (VEV). If an additional singlet scalar VEV $(u)$ gives rise to $U(1)_D$ gauge boson mass $M_{Z'}= g' u$ and also breaks the $U(1)_D$ spontaneously down to a remnant $Z_2$ symmetry under which $\chi_{1,2}$ are odd while all other fields are even, then the stability of $\chi_{1,2}$ is ensured, thus making them the viable DM candidates. Although the heavier DM can decay into the lighter one via singlet scalar coupling, we consider such off-diagonal Yukawa couplings to be negligible.
		
		For desired DM phenomenology, $\chi_{1}$ to be slightly heavier than $\chi_{2}$ with a mass splitting $\Delta m = {\cal O}$(100 keV) so that the former can annihilate into the latter with a cross-section: $\sigma (\chi_1 \chi_1 \to \chi_2 \chi_2)\approx 10^{-31} {\rm cm}^2$, providing a necessary flux of boosted $\chi_2$ to explain XENON1T excess. Note that such a large value of $\sigma (\chi_1 \chi_1 \to \chi_2 \chi_2)$ can be achieved through annihilation of $\chi_1$ to $\chi_2$ via $\Phi_2$ exchange. We further assume that $y_{1,2} \ll 1$, so that $\chi_1$ and $\chi_2$ abundances 
		can be generated at a later epoch, after they freeze-out from thermal bath, via the decay of $\Phi_1$.


		\section{Self-interaction of Dark Matter}
		\label{self_int}
		The dark sector particles have elastic self-scattering through $Z'$-mediated t-channel processes, thanks to the presence of terms like $g' Z'_{\mu}\overline{\chi_i}\gamma^\mu\chi_i$ in the model Lagrangian given by Eq.~\eqref{Lagrangian}. As we will see later, both $\chi_1$ and $\chi_2$ contribute to the present relic abundance of DM. Since their masses are very close to each other to give rise to the required boost factor and both have same gauge interactions, they contribute almost equally to the present DM abundance. Therefore, it suffices to discuss their self-interactions considering it to be a single component DM only. In order to explain small-scale astrophysical observations, the typical DM elastic scattering cross-section should be $\sigma \sim 1 \; {\rm cm}^2 (\frac{m_{\rm DM}}{g}) \approx 2\times 10^{-24} \; {\rm cm}^2 (\frac{m_{\rm DM}}{\rm GeV})$, which is many orders of magnitude larger than the typical weak-scale cross-section ($\sigma \sim 10^{-36} \; {\rm cm}^2$), suggesting the existence of a dark mediator much lighter than weak scale for DM mass around the electroweak ballpark. So we consider the $U(1)_D$ gauge boson of our model to be much lighter (order of magnitude lighter) than DM so that the non-relativistic DM scattering can be described by a Yukawa potential,
		\begin{equation}
			V(r)= \pm \frac{\alpha^\prime}{r}e^{-M_{Z'}r}
		\end{equation} 
		where the + (-) sign denotes repulsive (attractive) potential and $\alpha^\prime = g'^2/4\pi$ is the dark fine structure constant. While $\chi_i \overline{\chi_i}$ interaction is attractive, $\chi_i \chi_i$ and $\overline{\chi_i} ~\overline{\chi_i}$ are repulsive. We consider nearly degenerate masses for $\chi_1$ and $\chi_2$, hence $m_{\chi_1} \approx m_{\chi_2} = m_{DM}$.  To capture the relevant physics of forward scattering divergence for the self-interaction we define the transfer cross-section $\sigma_T$ as~\cite{Feng:2009hw,Tulin:2013teo,Tulin:2017ara}:
		\begin{equation}
			\sigma_T = \int d\Omega (1-\cos\theta) \frac{d\sigma}{d\Omega}
		\end{equation}
		
		In the Born Limit ($\alpha^\prime m_{DM}/M_{Z'}<< 1$), for both attractive as well as repulsive potentials, the transfer cross-section is: 
		\begin{equation}
			\sigma^{\rm Born}_T = \frac{8 \pi {\alpha^\prime}^2}{m^2_{DM} v^4} \Bigg(\ln(1+ m^2_{DM} v^2/M^2_{Z'} ) - \frac{m^2_{DM} v^2}{M^2_{Z'}+ m^2_{DM} v^2}\Bigg)\,.
		\end{equation} 
		Outside the Born regime ($\alpha^\prime m_{DM} /M_{Z'} \gtrsim 1 $), we have two distinct regions. In the classical limit ($m_{DM} v/M_{Z'}\gtrsim 1$), the solutions for an attractive potential is given by\cite{Tulin:2013teo,Tulin:2012wi,Khrapak:2003kjw}:
		\begin{equation}
			\sigma^{\rm classical}_T (\rm attractive)=\left\{
			\begin{array}{l}
				
				\frac{4\pi}{M^2_{Z'}}\beta^2 \ln(1+\beta^{-1}) ~~~~~~~~~~~~~~\beta \lesssim 10^{-1}\\
				\frac{8\pi}{M^2_{Z'}}{\beta^2}/{(1+1.5\beta^{1.65})} ~~~~~~~~~~~10^{-1}\lesssim \beta \lesssim 10^{3}\\
				\frac{\pi}{M^2_{Z'}}( \ln\beta + 1 -\frac{1}{2}\ln^{-1}\beta) ~~~~~~~~\beta \gtrsim 10^{3}\,,\\
			\end{array}
			\right.
		\end{equation}  
		and for the repulsive case;
		\begin{equation}
			\sigma^{\rm classical}_T (\rm repulsive)=\left\{
			\begin{array}{l}
				
				\frac{2\pi}{M^2_{Z'}}\beta^2 \ln(1+\beta^{-2}) ~~~~~~~~~~~~~~\beta \lesssim 1\\
				\frac{\pi}{M^2_{Z'}}(\ln 2\beta^2 -\ln ~\ln 2\beta)^2 ~~~~~~~~~~~ \beta \gtrsim 1\\
			\end{array}
			\right.
		\end{equation}  
		where $\beta = 2\alpha^\prime M_{Z'}/(m_{DM} v^2)$.
		\begin{figure}[h!]
			\includegraphics[scale=0.4]{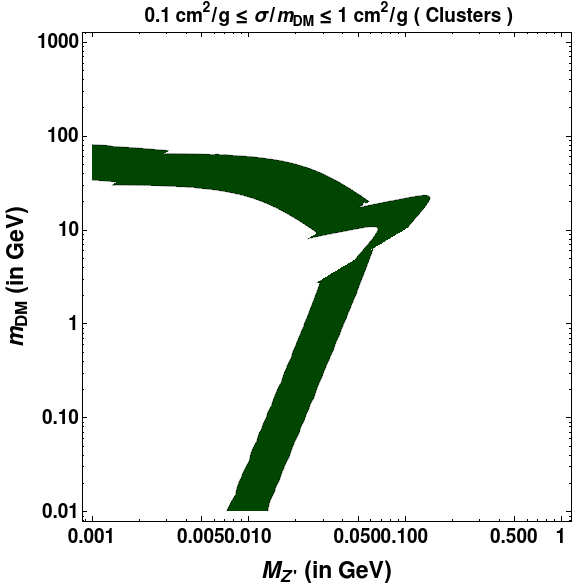}
			\hfil
			\includegraphics[scale=0.4]{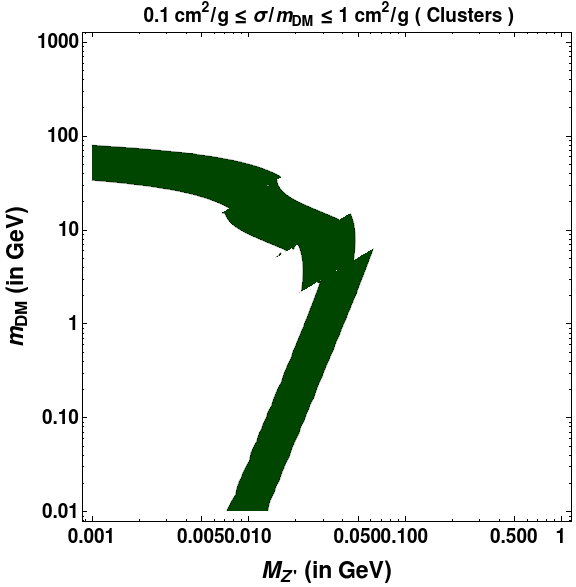}
			\caption{{Parameter space giving rise to attractive (left panel) and  repulsive (right panel) self-interaction cross-section in the range $0.1-1 \; {\rm cm}^2/{\rm g}$ for clusters ($v\sim1000 \; {\rm km/s}$).}}
			\label{sidm1}
		\end{figure}
		\begin{figure}[h!]	
			\includegraphics[scale=0.4]{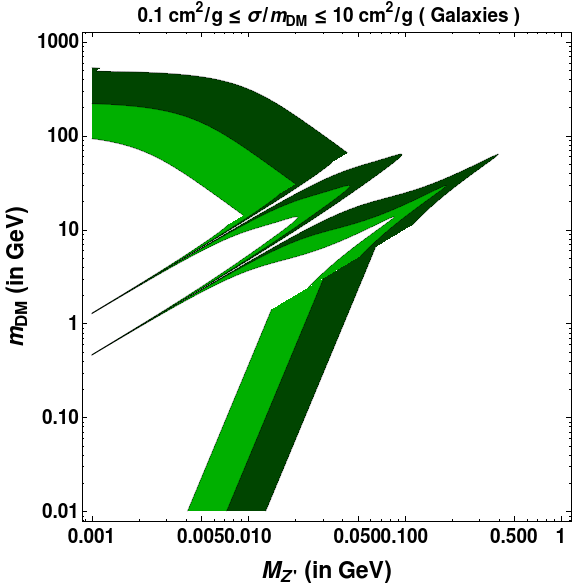}
			\hfil
			\includegraphics[scale=0.4]{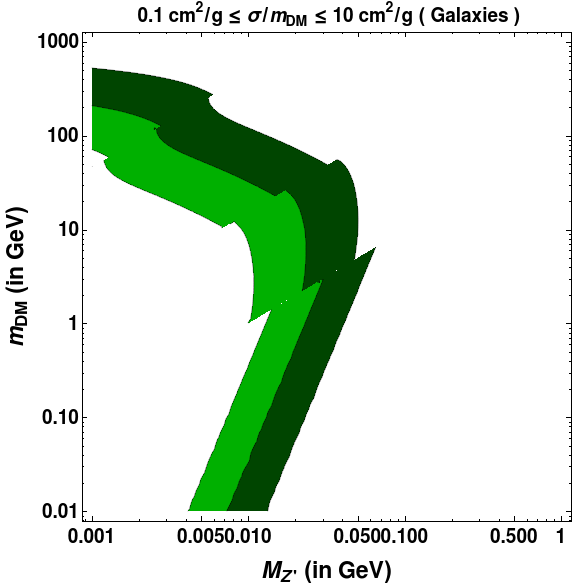}
			\caption{{Parameter space giving rise to attractive (left panel) and  repulsive (right panel) self-interaction cross-section in the range  $0.1-10 \; {\rm cm}^2/{\rm g}$ for galaxies ($v\sim200 \; {\rm km/s}$). Green color represents regions of parameter space where $1 \; {\rm cm}^2/{\rm g} < \sigma/m_{\rm DM}<10 \; {\rm cm}^2/{\rm g}$; Dark green colour  represents regions of parameter space where $0.1 \; {\rm cm}^2/{\rm g} < \sigma/m_{\rm DM} <1 \;{\rm cm}^2/{\rm g}$.}}
			\label{sidm2}
		\end{figure}
		\begin{figure}[h!]	
			\includegraphics[scale=0.4]{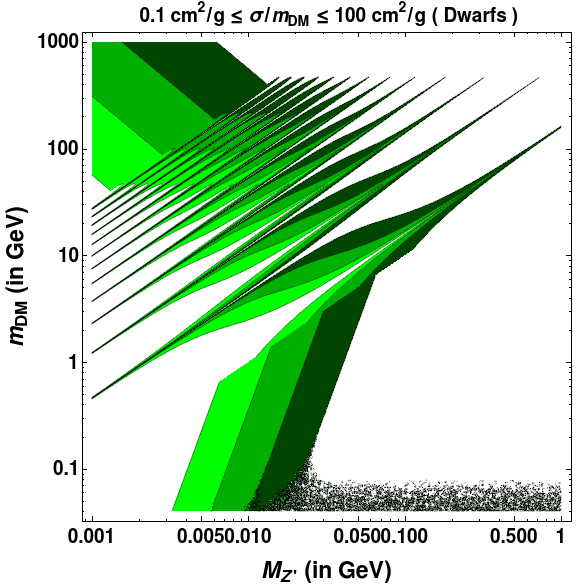}
			\hfil
			\includegraphics[scale=0.4]{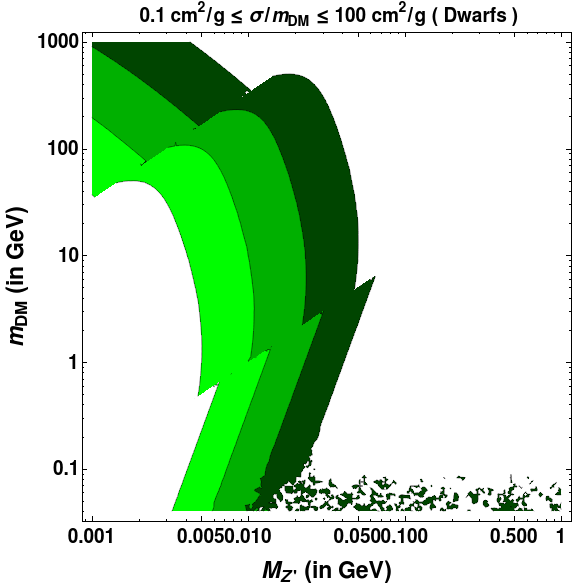}
			\caption{{Parameter space giving rise to attractive (left panel) and  repulsive (right panel) self-interaction cross-section $0.1-100 \; {\rm cm}^2/{\rm g}$ for dwarfs ($v\sim10 \; {\rm km/s}$). Lime green colour represents regions of parameter space where $10 \; {\rm cm}^2/{\rm g} < \sigma/m_{\rm DM} < 100 \; {\rm cm}^2/{\rm g}$ ;  Green colour represents regions of parameter space where $1 \; {\rm cm}^2/{\rm g} < \sigma/m_{\rm DM} <10 \; {\rm cm}^2/{\rm g}$; Dark green colour  represents regions of parameter space where $0.1 \; {\rm cm}^2/{\rm g} < \sigma/m_{\rm DM} <1 \;{\rm cm}^2/{\rm g}$.}}
			\label{sidm3}
		\end{figure}

		Outside the classical regime ($\alpha^\prime m_{DM}/M_{Z'} \gtrsim 1, m_{DM} v/M_{Z'} \lesssim 1$), we get the resonant regime where the cross-section is largely dominated by s-wave scattering. Here quantum mechanical resonances appear in $\sigma_T$ corresponding to (quasi-)bound states in the potential. In this regime, an analytical formula for $\sigma_T$ does not exist, and one has to solve the Schroedinger equation by partial wave analysis. Here we use the non-perturbative results for s-wave (l=0) scattering within the resonant regime obtained by approximating the Yukawa potential to be a Hulthen potential $\Big( V(r) = \pm \frac{\alpha^\prime \delta e^{-\delta r}}{1-e^{-\delta r}}\Big)$ which is given by~\cite{Tulin:2013teo}:
		\begin{equation}
			\sigma^{\rm Hulthen}_T = \frac{16 \pi \sin^2\delta_0}{m^2_{DM} v^2}
		\end{equation}
		where l=0 phase shift is given in terms of the $\Gamma$ functions as :
		\begin{equation}
			\delta_0 =arg \Bigg(\frac{i\Gamma \Big(\frac{i m_{DM} v}{k M_{Z'}}\Big)}{\Gamma (\lambda_+)\Gamma (\lambda_-)}\Bigg),~~~~ \lambda_{\pm} = \left\{
			\begin{array}{l}
				1+ \frac{i m_{DM} v}{2 k M_{Z'}}  \pm \sqrt{\frac{\alpha^\prime m_{DM}}{k M_{Z'}} - \frac{ m^2_{DM} v^2}{4 k^2 M^2_{Z'}}} ~~~~ {\rm Attractive}\\
				1+ \frac{i m_{DM} v}{2 k M_{Z'}}  \pm i\sqrt{\frac{\alpha^\prime m_{DM}}{k M_{Z'}} + \frac{ m^2_{DM} v^2}{4 k^2 M^2_{Z'}}} ~~~~ {\rm Repulsive}\\
			\end{array}
			\right.
		\end{equation}   
		and $k \approx 1.6$ is a dimensionless number. The differential cross-section is $d\sigma/d\Omega = \sigma_T /(4 \pi)$. 
		
		Using these self-interaction cross-sections and using the required $\sigma/m_{\rm DM}$ from astrophysical observations at different scales, we constrain the parameter space of the model in terms of DM $(\chi_{1,2})$ and mediator $Z'$ masses. In Fig.~\ref{sidm1},\ref{sidm2},\ref{sidm3}, keeping $g'$ fixed at $0.1$, we show the allowed parameter space in DM mass versus $Z'$ mass plane which gives rise to the required DM self-interaction cross-section ($\sigma/m_{\rm DM}$) in the range $\sigma \in 0.1-1~{\rm cm}^2/{\rm g}$ for clusters ($v\sim1000~ \rm km/s$), $\sigma \in 0.1-10~{\rm cm}^2/{\rm g}$ for galaxies ($v\sim 200~ \rm km/s$)and $\sigma \in 0.1-100~{\rm cm}^2/{\rm g}$ dwarf galaxies ($v\sim 10~ \rm km/s$) respectively. Because of the light vector mediator, here we can have both attractive and repulsive interactions, unlike in the case with a scalar mediator where the interactions are purely attractive. The sharp spikes in the left panels of Fig.~\ref{sidm2},\ref{sidm3} are the patterns of quantum
		mechanical resonances and anti-resonances for the attractive potential case, which is absent for the repulsive
		case, shown on the right panels. It is clear that the resonant regime
		corresponds to a large region of parameter space. These features are more prominent for the galactic and dwarf galactic scales where DM has smaller velocities. This is due to the fact that for a fixed $\alpha^\prime$, the condition $m_{DM} v/M_{Z'} < 1$ governs the onset of
		quantum mechanical and non-perturbative effects. Clearly, a wide range of DM mass is allowed from the self-interaction requirements, but mediator mass is constrained within one or two orders of magnitudes (except in the resonance regimes) from both cosmological and astrophysical requirements. We will finally compare these regions of sub-GeV scale DM mass parameter space in the context of XENON1T excess and other phenomenological constraints.
		
\begin{figure}
			$$
			\includegraphics[scale=0.35]{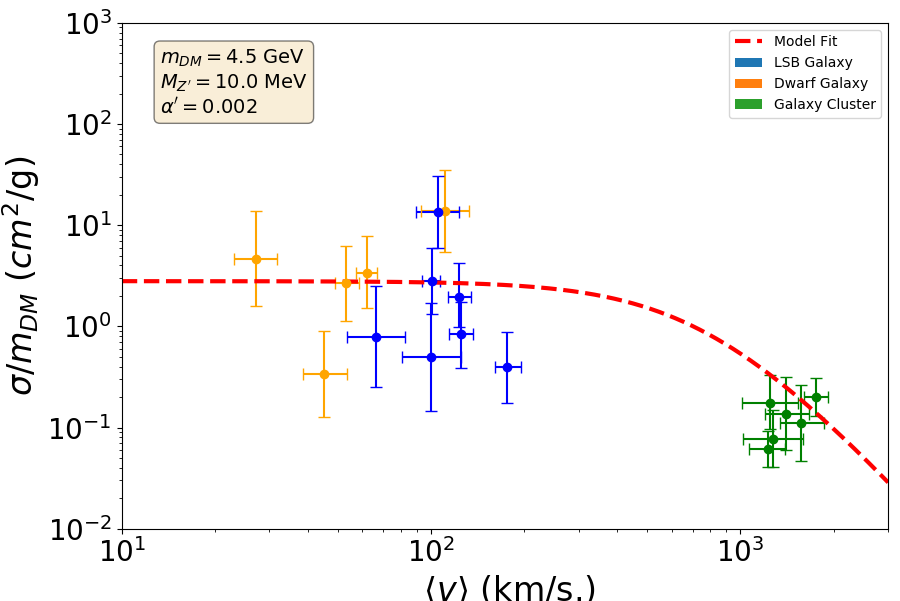}
			$$
			\caption{{The self-interaction cross section per unit mass of DM as a function of average collision velocity.}}
			\label{astrofit}
		\end{figure}

		The self-interaction cross-section per unit DM mass as a function of average collision velocity is shown in figure~\ref{astrofit} as measured from astrophysical data. The data includes measurements from dwarfs (orange), LSBs (blue) and clusters (green)~\cite{Kaplinghat:2015aga,Kamada:2020buc}. The red dashed curve corresponds to the velocity-dependent cross-section calculated from our model for a particular set of benchmark values (i.e $m_{\rm DM}=4.5~\rm GeV$, $M_{Z'}=10~\rm MeV$ and $\alpha^\prime = 0.002$) allowed from all relevant phenomenological constraints. It is evident from Fig.~\ref{astrofit} that the scenario discussed in this work explains the astrophysical observation of velocity-dependent DM self-interaction remarkably well.
		
		\section{Relic Density of Dark Matter}
		\label{dm_production}
\begin{figure}[ht]
			\includegraphics[scale=0.25]{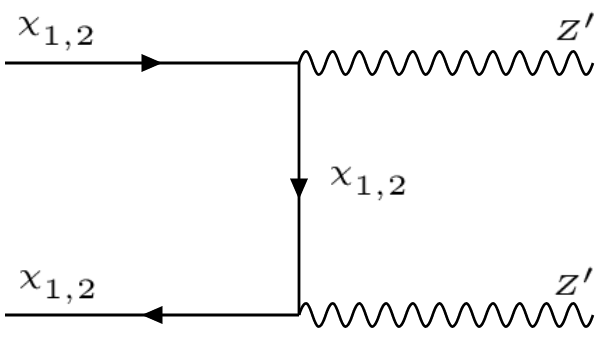} 
			\hfil
			\includegraphics[scale=0.25]{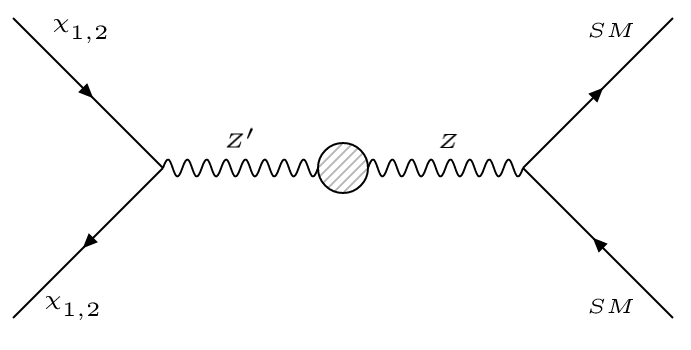}
			\caption{{Feynman diagrams for dominant number changing processes of DM.}}
			\label{thermal_relic}
		\end{figure}

\begin{figure}
			\centering
			\includegraphics[scale=0.45]{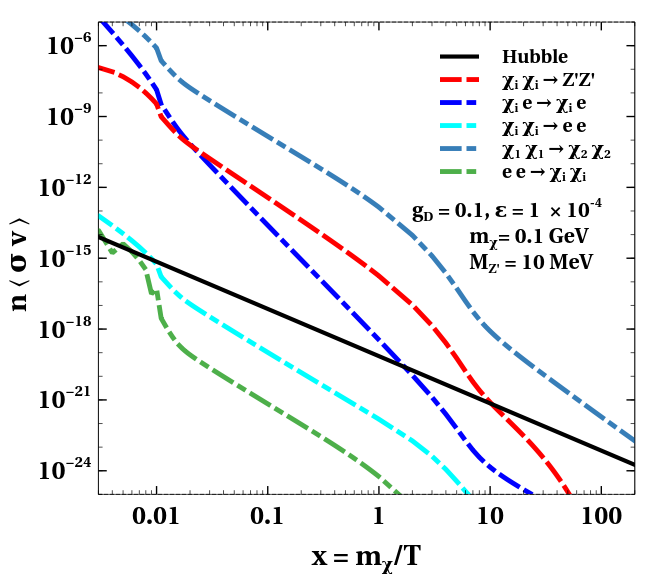}
			\caption{{Comparison of different scattering processes involving DM with Hubble rate of expansion.}}
			\label{Decoupling}
		\end{figure}

		There exists several frameworks for production of SIDM in the literature  \cite{Kouvaris:2014uoa, Bernal:2015ova, Kainulainen:2015sva, Hambye:2019tjt, Cirelli:2016rnw, Kahlhoefer:2017umn, Belanger:2011ww}. We adopt a minimalistic approach here by first considering the usual $2 \leftrightarrow 2$ vector portal interactions between DM and SM sectors. As DM can interact with itself via both $Z'$ and singlet scalar interactions, the vector portal always remains  dominant due to light $Z'$ and sizeable $g'$. On the other hand, DM can interact with the SM bath only via kinetic mixing of neutral vector bosons or singlet scalar mixing with the SM Higgs boson. However, we ignore the DM-SM interaction via scalar portal in this work and try to constrain the gauge portal maximally from all relevant phenomenology. The dominant number changing processes for DM are the ones shown in Fig.~\ref{thermal_relic}. DM-SM interaction via sizeable kinetic mixing ($\sim 10^{-4}$) is responsible for bringing the dark sector to thermal equilibrium in the early Universe. To check whether DM-SM interactions can reach equilibrium in the early Universe, we compare in Fig.~\ref{Decoupling}, the rates of different annihilation processes considering {\color{red}$\epsilon = 10^{-4}$} with the Hubble expansion rate of the universe in a radiation dominated era. For numerical analysis, the model has been implemented in \texttt{LanHEP} \cite{Semenov:2014rea} and \texttt{CalcHEP} \cite{Belyaev:2012qa}. 
		As we can see from Fig.~\ref{Decoupling}, rate of processes like $\chi e \to \chi e$ are well above the Hubble expansion rate at early epochs keeping DM $\chi$ in thermal equilibrium. However, to get velocity dependent self-scattering we are considering heavier DM compared to the mediator {\it i.e.,} $m_{\chi} > M_{Z'}$ and therefore, DM has large annihilation cross section to $Z'$ pairs compared to its annihilation rates into SM particles, the later being kinetic mixing suppressed. This can significantly lead to suppressed thermal relic of DM. The dominant number changing processes contributing to its thermal freeze-out are shown in Fig.~\ref{thermal_relic}. 
		The thermally averaged cross-section for the t-channel process $\chi_i \chi_i \rightarrow Z' Z'$ shown in the left panel of Fig.~\ref{thermal_relic} is
		\begin{equation}
			\langle\sigma v\rangle \sim \frac{\pi {\alpha^\prime}^2}{m^2_{\rm DM}}
		\end{equation}
		where $m_{\rm DM}$ denotes the masses of $\chi_{1,2}$ which are very close to each other. For typical gauge coupling and DM mass of our interest namely, $\alpha^\prime \sim 0.001, m_{\rm DM} \sim 0.1$ GeV, this leads to a cross-section which is at least two orders of magnitudes larger compared to the typical annihilation cross-section of thermal DM. Thus it reduces the relic abundance by the same order of magnitudes.


		Since we have two singlet fermions $\chi_{1,2}$ with tiny mass difference, identical gauge couplings and a strong $\chi_1 \chi_1 \rightarrow \chi_2 \chi_2$ conversion rate via $\Phi_2$ exchange (required for boosted DM phenomenology to be discussed in an upcoming section), we need to solve relevant Boltzmann equations for both of them. Additionally, as thermal relic of both $\chi_{1,2}$ will be sub-dominant due to large annihilation rates into $Z'$ pairs, we consider an additional singlet scalar $\Phi_1$ whose late decay can fill this deficit. Therefore, for a complete numerical analysis of DM relic abundance, we need to solve three coupled Boltzmann equations for $\chi_{1,2}$ and $\Phi_1$. Unlike $\chi_{1,2}$ whose interactions with the SM bath are suppressed due to small kinetic mixing, the scalar singlet can be in thermal equilibrium with the SM due to large quartic couplings leading to thermal freeze-out followed by late decay into DM\footnote{Similar hybrid setup can also be found in earlier works, for example, \cite{Feng:2003uy, Borah:2019bdi, Biswas:2018ybc, Borah:2018gjk, Borah:2017dfn}.}. Defining the comoving number densities of these particles as $Y_{\chi_{1,2}}=n_{\chi_{1,2}}/s(T), Y_{\Phi_1} = n_{\Phi_1}/s(T)$, the relevant coupled Boltzmann equations can be written as follows.		
		\begin{equation}
			\begin{aligned}
				&\frac{dY_{\Phi_1}}{dx}= -\frac{s(m_{\rm DM})}{x^2  H(m_{\rm DM})} \langle\sigma(\Phi_1 \Phi_1 \to {\rm SM \; SM}) v \rangle \Bigg(Y^2_{\Phi_1}-\big(Y^{\rm eq}_{\Phi_1}\big)^2\Bigg) \\&-\frac{x}{H(m_{\rm DM})}\Bigg(\langle \Gamma_{\Phi_1 \rightarrow \chi_1 \chi_1}\rangle + \langle\Gamma_{\Phi_1 \rightarrow \chi_2 \chi_2}\rangle \Bigg) Y_{\Phi_1} \hspace{0.0cm};
				\\&
				\frac{dY_{\chi_1}}{dx}=-\frac{s(m_{\rm DM})}{x^2  H(m_{\rm DM})} \Bigg(\langle\sigma(\chi_1 \chi_1  \to Z' Z') v\rangle \Big(Y^2_{\chi_1} - (Y^{eq}_{\chi_1})^2\Big) \\& + \langle \sigma(\chi_1 \chi_1 \to \chi_2 \chi_2) v \rangle \Big(Y^2_{\chi_1} - \frac{(Y^{\rm eq}_{\chi_1})^2}{(Y^{\rm eq}_{\chi_2})^2}Y_{\chi_2}^2\Big)\Bigg)
				+ \frac{x}{H(m_{\rm DM})} \langle \Gamma_{\Phi_1 \rightarrow \chi_1 \chi_1}\rangle Y_{\Phi_1}\hspace{0.0cm};
				\\&
				\frac{dY_{\chi_2}}{dx}= - \frac{s(m_{\rm DM})}{x^2  H(m_{\rm DM})}\Bigg( \langle\sigma(\chi_2 \chi_2  \to Z' Z') v\rangle \Big(Y^2_{\chi_2} -(Y^{eq}_{\chi_2})^2\Big) \\& - \langle \sigma(\chi_1 \chi_1 \to \chi_2 \chi_2) v\rangle \Big(Y^2_{\chi_1} - \frac{(Y^{\rm eq}_{\chi_1})^2}{(Y^{\rm eq}_{\chi_2})^2}Y_{\chi_2}^2\Big)\Bigg)
				+ \frac{x}{H(m_{DM})} \langle \Gamma_{\Phi_1\rightarrow \chi_2 \chi_2}\rangle Y_{\Phi_1}
			\end{aligned}
		\end{equation}
		
		where, $x=\frac{m_{\rm DM}}{T}$, $s(m_{\rm DM})= \frac{2\pi^2}{45}g_{*S}m^3_{\rm DM}$ , $H(m_{\rm DM})=1.67 g^{1/2}_*\frac{m^2_{\rm DM}}{M_{\rm Pl}}$ and $\langle \sigma(\Phi_1 \Phi_1\to {\rm SM \;SM}) v\rangle$ represents the thermally averaged cross-section~\cite{Gondolo:1990dk} of annihilation of $\Phi_1$ to all SM particles. The relevant cross-sections and decay widths are given in appendix \ref{appen2}. Also, as mentioned earlier, $m_{\rm DM} = m_{\chi_1} \approx m_{\chi_2}$. 
		Note that the total $\Phi_1$ decay width ($\Gamma_{\Phi_1}$) is assumed to be very small, leading to conversion of $\Phi_1$ into DM at a late epoch, nevertheless well before the big bang nucleosynthesis (BBN). In fact, the chosen decay ($\Gamma_{\Phi_1}=8.8 \times 10^{-23}$ GeV) corresponds to a lifetime of approximately $6.4 \times 10^{-3}$ s. 
		\begin{figure}[h!]
			\centering
			\includegraphics[scale=0.5]{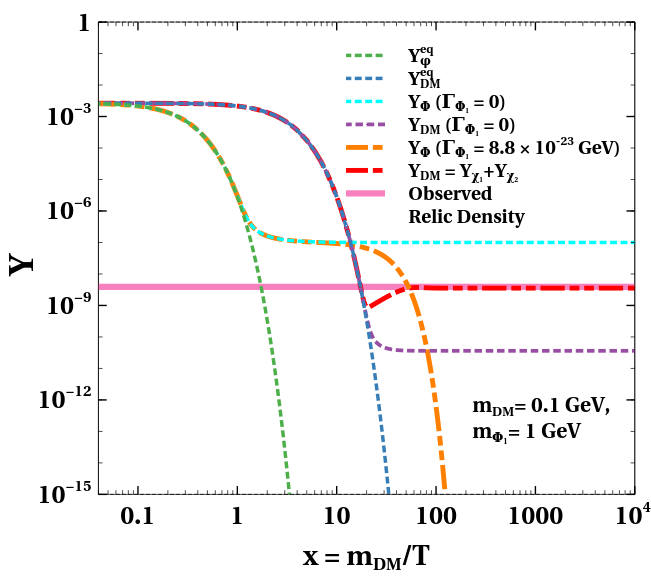}
			\caption{Comoving number densities of dark sector particles considering different sub-processes indicated in the legends.}
			\label{relic}
		\end{figure} 
		
		The evolution of these comoving number densities are shown in Fig. \ref{relic}. To understand the importance of different processes in the Boltzmann equations, we show DM generation incorporating different annihilation or decay processes separately. The equilibrium number density of DM and the scalar $\Phi_1$ are shown in dashed blue and green curves. The thermal freeze-out abundances of DM and scalar $\Phi_1$ are shown in dashed violet and cyan curves. The thermal freeze-out abundance of DM is clearly well below the  observed relic density (shown by the magenta line).
		On the other hand, the singlet scalar $\Phi_1$ freezes out from the bath leaving a sizeable relic. At late epochs, the scalar singlet decays into DM, filling the DM relic deficit as shown by the red dot-dashed line. The singlet scalar abundance including its late decay is shown by the orange dot-dashed line. Since both the singlet fermions $\chi_{1,2}$ have the same gauge coupling and tiny mass splitting, they get generated in almost equal amount from the bath and consequently from dark freeze-out. 
		$\Phi_1$ decays to both $\chi_{1,2}$ equally and the interconversion between the two components does not affect the final relic significantly due to identical gauge interactions and tiny mass splitting between them, hence both $\chi_{1,2}$ are almost equally abundant. This is in sharp contrast with other boosted DM scenarios, for example \cite{Borah:2021jzu}, where different final abundances of two DM fermions were found due to their different gauge interactions. Note that the decay of $\Phi_2$ does not significantly affect the relic since the Yukawa coupling $y'$ (see model Lagrangian~\ref{Lagrangian}) is large and hence $\Phi_2$ decays to $\chi_{1,2}$ much before $\chi_{1,2}$ undergoes dark freeze-out. Any such initial abundance of DM will eventually get diluted due to strong annihilation into dark gauge boson ($Z'$) and the correct relic can only be obtained with late decay of $\Phi_1$.
		
		\newcommand{\eqdef}{\overset{\mathrm{def}}{=\joinrel=}}
		


		\section{Boosted Dark Matter and Xenon1T excess}
		\label{xenon}
\begin{figure}[h!]
			\centering
			\includegraphics[scale=0.5]{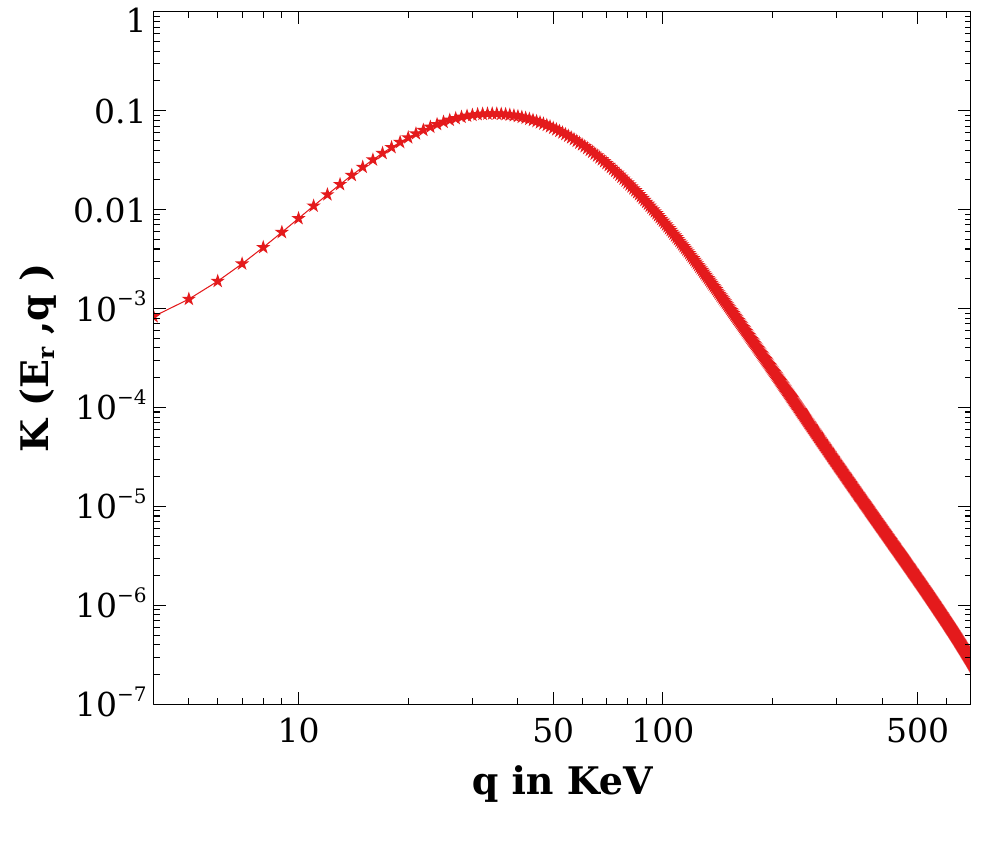}
			\caption{\footnotesize{ Atomic excitation factor is shown as a function of momentum transferred.}}
			\label{aef}
		\end{figure}

		The DM interpretation of the XENON1T excess with conventional dark matter is not possible, essentially because
		of its non-relativistic nature. For DM sufficiently heavier than the electron, the electron recoil (kinetic) energy lies in a range of $\mathcal{O}(\rm eV)$~({\it i.e.} $E_r\sim  m_e \times (10^{-3}
		c)^2 \simeq \mathcal{O}(\rm eV)$ where $m_e$ is mass of electron and $v\sim 10^{-3}c$ being
		the typical velocity of cold dark matter). On the other hand, XENON1T collaboration has reported an excess of electron recoil events over the background in the recoil energy $E_r$ in a range 1-7 keV, 
		peaked around 2.4 keV \cite{XENON:2020rca}. This essentially implies that the energy deposition by conventional non-relativistic DM can not explain the excessive events of $\mathcal{O}(keV)$ as reported by the XENON1T collaboration.
		However, in scenarios involving a mechanism to exert sufficient boost onto a DM component, it is possible to explain the XENON1T excess through the elastic scattering of the boosted DM component off electron at the XENON1T detector. In our setup, at the present day, non-relativistic DM particle $\chi_1$ particles annihilates in the galactic center, producing boosted final state particles $\chi_2$, with Lorentz boost factor $\gamma = m_{\chi_1}/m_{\chi_2}$.
		For a fixed incoming velocity $v$ of DM fermion, the differential 
		scattering cross-section for the elastic scattering process $\chi_2 e \rightarrow 
		\chi_2 e$ can be written as
		\begin{equation}
			\frac{d \langle \sigma v \rangle }{d E_r} = \frac{\sigma_e}{2 m_e v } \int_{q-}^{q+} a^2_0~ q~ dq~|F(q)|^2~ K(E_r,q)\,,
			\label{Event_central}
		\end{equation}	
		where $m_e$ is the electron mass, $\sigma_e$ is the corresponding free electron cross section at fixed momentum transfer 
		$q=1/a_0$ with $a_0 = \frac{1}{\alpha m_e}$ being the Bohr radius, $\alpha = \frac{e^2}{4 \pi}=\frac{1}{137}$ being 
		the fine structure constant, $E_r$ is the recoil energy of electron and $K(E_r, q)$ is the atomic excitation factor. For our calculations, we adopt the the atomic excitation factor from \cite{Roberts:2019chv} and we assume the DM fermion form factor to be unity. The variation of atomic excitation factor with the transferred momentum $q$ is shown in Fig.~\ref{aef}. Here, the dominant contribution comes from the bound states with principal quantum number $n=3$ as their binding energy is around a few keVs.

		From the kinematics of the elastic scattering, the limits of integration for Eq.~\eqref{Event_central} are given by
\begin{equation}
			q_\pm=m_{\chi_{2}} v \pm \sqrt{m^2_{\chi{_2}} v^2 -2m_{\chi_{2}}E_r}\,.
		\end{equation}
		The differential event rate for the scattering of $\chi_2$ with electrons in Xenon atom at XENON1T detector, {\it i.e} $\chi_2 e 
		\rightarrow \chi_2 e$, can then be written as 
		\begin{equation}
				\frac{dR}{dE_r}=n_T ~\Phi_{\chi_{2}} \frac{d \langle \sigma v \rangle}{d E_r}
				\label{event_rate}
			\end{equation}
		 Here  $n_T= {4\times10^{27} ~\rm Ton}^{-1}$ is the number of target atoms and $\Phi_{\chi_2}$ is the flux of the boosted $\chi_2$ particle.
		The final detected recoil energy spectrum can be obtained by convolving Eq.~\eqref{event_rate} with the energy resolution of the XENON1T 
		detector. The energy resolution of the detector is given by a Gaussian distribution with an 
		energy dependent width, 
		\begin{equation}
			\zeta(E,E_r)=\frac{1}{\sqrt{2 \pi \sigma^2_{\rm det}}}{\rm Exp}\Big[-\frac{(E-E_r)^2}{2 \sigma^2_{\rm det}}\Big] \times \gamma(E)
		\end{equation}
		where $\gamma(E)$ is detector efficiency reported in Fig.~2 of \cite{Aprile:2020tmw} and the width $\sigma_{\rm det}$ is given by 
		\begin{equation}
			\sigma_{\rm det}(E)= a \sqrt{E} + b E
		\end{equation} 
		with $a=0.3171$ and $b=0.0037$.
		
		In this boosted DM approach to explain the XENON1T excess, DM $\chi_1$ which contributes to half of the total DM density in the present universe annihilates into dark matter $\chi_2$ giving a significant boost to explain the reported excess. In the present universe $\chi_1$ can be assumed to annihilate to $\chi_2$ only in DM dense regions like the Galactic center (GC) or the Sun\footnote{In the case of boosted flux from the Sun, strong evaporation bound \cite{Griest:1986yu, Gould:1987ju} forces us to choose DM mass in the GeV regime where DM-nucleon scattering rate faces tight constraints from direct search experiments like CRESST-III~\cite{CRESST:2019jnq}. Thus, the required $\chi_2$ flux from solar captured $\chi_1$ can not be obtained.}.  
		
		The flux of $\chi_{2}$ from GC is given by~\cite{Agashe:2014yua}:
		\begin{align}
			\frac{d\Phi^{\rm GC}_{\chi_2}}{d \Omega dE_{\chi_2}}=\frac{r_{Sun}}{16 \pi} \left(\frac{\rho^{\rm DM}_{\rm local}}{m_{\chi_{1}}}\right)^{2} \mathcal {J} ~~\langle \sigma_{ \chi_1\chi_1\rightarrow\chi_2\chi_2}v \rangle \frac{d n_{\chi_2}}{dE_{\chi_2}}
		\end{align}
		where $r_{Sun}$ is the distance from the sun to the GC ($r_{Sun}=8.33 kpc$) and $\rho^{\rm DM}_{\rm local}$ is the local DM density. As two mono-energetic $\chi_2$ particles with energy $m_{\chi_1}$ are produced by the $\chi_1\chi_1\rightarrow\chi_2\chi_2$ annihilation process, so the differential energy spectrum can be written as:
		\begin{equation}
			\frac{dn_{\chi_2}}{dE_{\chi_2}}= 2 \delta(E_{\chi_2}-m_{\chi_2})
		\end{equation} 
		and the halo-shape dependent dimension less quantity $\mathcal{J}$ is given by:
		\begin{equation}
			\mathcal{J}= \int_{l.o.s} \frac{ds}{dr_{Sun}}\left(\frac{\rho(r(s,\theta))}{\rho^{\rm DM}_{local}}\right)^2
		\end{equation}
		Here $r(s,\theta)=(r^2_{Sun}+s^2-2 r_{Sun} s \cos\theta)^{\frac{1}{2}}$ is the coordinate centered on the GC where $s$ is the line-of-sight distance to the earth and $\theta$ is the angle between the line-of-sight direction and the earth/GC axis.
		
		Assuming the DM follows a Navarro-Frenk-White profile~\cite{Navarro:1995iw}, and integrating over the whole sky, the obtained BDM flux is~\cite{Agashe:2014yua}:
		\begin{equation}
			\Phi^{\rm GC}_{\chi_2}= 1.68\times10^3 \; {\rm cm}^{-2} {\rm s}^{-1} \bigg(\frac{\langle\sigma_{ \chi_1\chi_1\rightarrow\chi_2\chi_2}v\rangle}{3.52\times10^{-31} \; {\rm cm}^2}\bigg)\bigg(\frac{0.1\; {\rm GeV}}{m_{\chi_{1}}}\bigg)^2
			\label{flux}
		\end{equation}
		It is worth mentioning here that, though the DM density peaks toward the GC, since XENON1T cannot distinguish the direction of the incoming DM particle, all sky directions should be included. Note that the $\langle\sigma_{ \chi_1\chi_1\rightarrow\chi_2\chi_2}v\rangle$ is 5 orders of magnitude larger than the typical WIMP annihilation cross-section, however as can be seen in the appendix~\ref{mw_dm_evolution}, it does not alter the abundance of $\chi_1$ and $\chi_2$ in the Milky Way over the galactic time scale nor does it affect the mass of the Milky Way due to DM evaporation.

		%
		
		\begin{figure}[h!]
			\centering
			\includegraphics[scale=0.35]{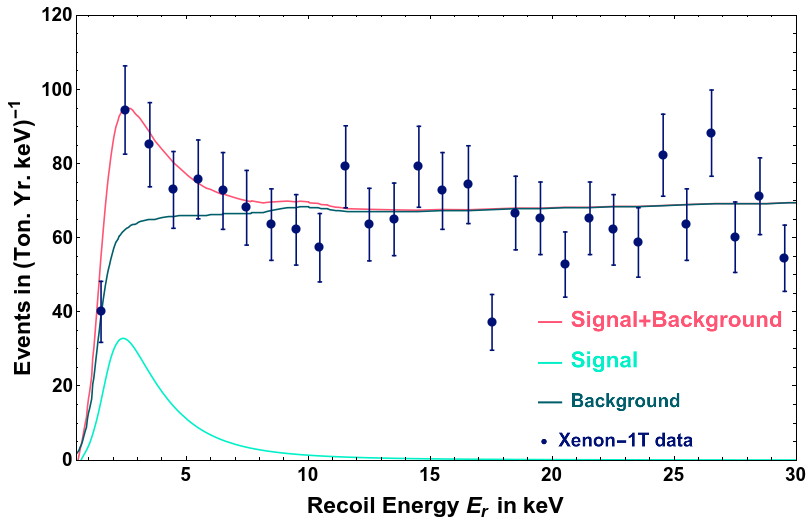}
			\caption{{Fit to XENON1T electron recoil excess with the Boosted dark matter}}
			\label{xenonfit}
		\end{figure}

		The free electron scattering cross-section for the process $\chi_2 e \rightarrow \chi_2 e$ is given by
		\begin{equation}
			\sigma_e = \frac{ g'^2 \epsilon^2 g^2  m^2_e  }{\pi M^4_{Z'}}
			\label{DM-electron-scattering}
		\end{equation}
		where $\epsilon$ is the kinetic mixing parameter between $Z$ and $Z'$ gauge bosons, $g$ is the weak gauge coupling and $g'$ is the $U(1)_D$ gauge coupling. As already mentioned, for DM sufficiently heavier than electron, the recoil cross-section $\sigma_e$ 
		is independent of DM mass as the reduced mass is almost equal to electron mass.

\begin{figure}[h!]
			\centering
			\includegraphics[scale=0.5]{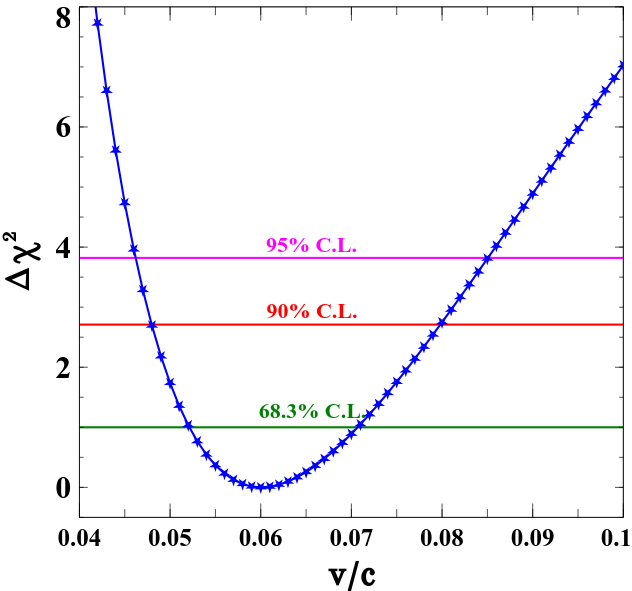}
			\caption{Shows the $\chi^2$ fit for velocity $v$ that gives the best fit and the allowed ranges of $v$ {\it i.e.}  $v \in [0.052,0.07]$($68\%$ C.L.) and $v \in [0.046,0.085]$($95\%$ C.L.)  }
			\label{chisqr}
		\end{figure}

		Thus the final detected recoil energy spectrum is given by
		\begin{equation}
				\frac{dR_{\rm det}}{dE_r}=\frac{n_T \Phi_{\chi_2} \sigma_e a^2_0}{2 m_e v}  \int dE ~~\zeta(E,E_r) \Bigg[\int_{q-}^{q+} dq~~ q~ K(E_r,q)\Bigg] 	
		\end{equation}
		With the flux mentioned in Eq.\eqref{flux}, the electron scattering cross-section $\sigma_e$ that can explain the electron recoil excess at XENON1T is calculated to be $6.3\times 10^{-9}$~${\rm GeV}^{-2}$. To obtain the fit to XENON1T data shown in Fig.~\ref{xenonfit} we have used benchmark values $m_{\chi_{2}}= 90$ MeV, $v=0.06$ which gives the best fit to the data. Such velocity can be obtained by fixing $\Delta m/m_{\chi_{2}} =2\times 10^{-3}$ where $\Delta m = m_{\chi_{1}} - m_{\chi_{2}}$ giving rise to the necessary boost factor. This is governed by the velocity mass-splitting relation given by:
		\begin{equation}
			v=\sqrt{1-\Big(1+\frac{\Delta m}{m_{\chi_2}}\Big)^{-2}}.
		\end{equation} In Fig.~\ref{chisqr}, we present the $\chi^2$-fit for velocity $v$ that gives the best fit and the allowed ranges of velocity for different confidence interval {\it i.e.}  $v \in [0.052,0.07]$($68\%$ C.L.) and $v \in [0.046,0.085]$($95\%$ C.L.). Thus it is worth mentioning here that the appropriate boost can be achieved by tuning the mass-splitting between the two DM components in a range $\Delta m/m_{\chi_{2}}=(1.25-4)\times10^{-3}$.
		
	\section{Direct Detection through DM-Nucleon scattering}\label{direct_search}
		
		Here it is worth mentioning that, even though the the DM direct detection experiments like CRESST-III and XENON1T which look for the DM-nucleon scattering signals, are not sensitive to such light DM, however, because of the larger velocity of the boosted DM, it has the potential to trigger DM-nucleon scattering as the typical momentum transfer in such a case is of the order $\mathcal{O}(10)$ MeV, for the parameter space we are interested in. Thus a light boosted DM will mimic an ambient non-relativistic DM particle having a mass $v_{BDM}/v_{CDM}$ times larger than the mass of the boosted DM ($m_{\chi_2}$) where $v_{BDM}$ is the velocity of the boosted DM and $v_{CDM}$ is the velocity of the so called '{\it vanilla}' dark matter which is equal to $0.001c$. Hence it is instructive to confront the model parameters against the constraints from these DM direct detection experiments. 
		
		\begin{figure}[h!]
			\centering
			\includegraphics[scale=0.6]{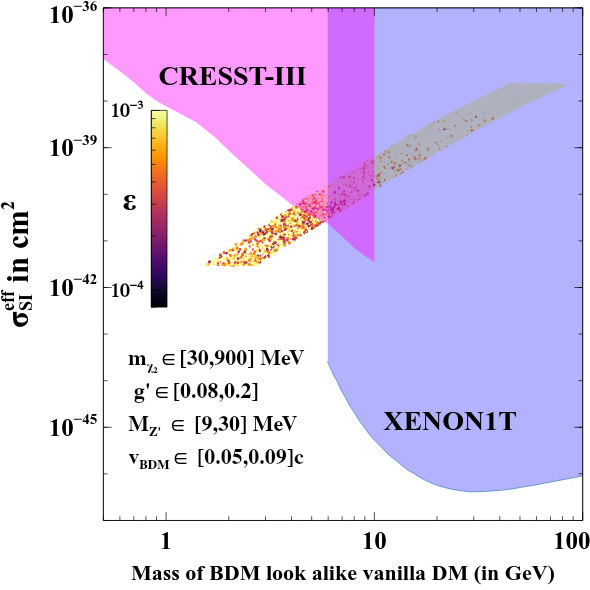}
			\caption{Spin-independent DM-nucleon scattering cross-section as a function of the BDM look-alike vanilla DM mass. The range of the parameters used in the scan are mentioned in the inset of the figure. The color code shows the value of the $\epsilon$ required to get the fit to XENON1T electron recoil excess.}
			\label{dd_con}
		\end{figure}
		
		The spin-independent elastic scattering cross-section of the DM with the nucleon is computed as:
			\begin{equation}
				\sigma_{SI}=\frac{g^2 g'^2 \epsilon^2 m^2_{DM} m^2_{N}}{\pi M^4_{Z'}(m_{DM}+m_{N})^2}
			\end{equation}
			where $m_{N}$ is the mass of the nucleon. However, to compare this cross-section in our model with the experimental constraints, we have to define the effective scattering cross-section $\sigma^{eff}_{SI}=\mathcal{R}~\sigma_{SI}$ where $\mathcal{R}$ is the ratio of the boosted DM flux to the ambient non-relativistic local cold DM flux. This is the algorithm that is followed in multi-component DM scenarios to apply direct search constraints on each DM components. This ratio $\mathcal{R}$ is of the order $\mathcal{O}(10^{-8})$ or smaller in our case depending on the DM mass. In Fig.~\ref{dd_con}, we have shown this effective DM-nucleon scattering cross-section as a function of the look alike '{\it vanilla}' DM mass for the corresponding boosted DM. We performed a scan for our model parameters as mentioned in the inset of Fig.~\ref{dd_con}, and scrutinized the points that satisfy the self-interaction criteria. Then the annihilation cross-section $\langle\sigma_{ \chi_1\chi_1\rightarrow\chi_2\chi_2}v\rangle$ of the DM is chosen such that it does not affect the $\chi_{1,2}$ abundance in Milky Way as discussed in Appendix~\ref{mw_dm_evolution}. With the obtained flux, then the DM-electron scattering cross-section is calculated such that it can give rise to the observed electron recoil excess and the required kinetic mixing $\epsilon$ is estimated as other parameters are already constrained from the self-interaction criteria. This is shown in the color code in Fig.~\ref{dd_con} and clearly there are various points which are safe from the constraints on $\epsilon$ from dark photon searches\cite{Bauer:2018onh}. And finally the effective scattering cross-section for the DM-nucleon scattering of the boosted DM is computed as discussed above. As we can see, some light DM mass is still safe from these constraints whereas the larger DM masses are ruled out by the stringent constraints from the XENON1T experiment.

		\section{Summary and Conclusion}
		\label{conclusion}
		We have proposed a boosted self-interacting dark matter scenario as a possible origin of XENON1T electron excess adopting a minimal setup where DM is composed of two vector-like singlet fermions charged under a dark abelian gauge symmetry. While sufficient DM self-interactions can be generated due to the existence of a light vector boson, the XENON1T excess can be realized from the boosted component of DM scattering off electrons. A sufficient boost factor can be realized by tuning the mass splitting between two DM fermions and the cross-section of their inter-conversion. While DM can be produced from the thermal bath via freeze-in mechanism due to tiny kinetic mixing of neutral vector bosons, the final abundance remains suppressed due to large DM annihilation rates within the dark sector. The deficit can be filled through the late decay of a singlet scalar which freezes out earlier from the thermal bath. Adopting suitable benchmark values, we have shown how correct relic of DM can be generated by solving the coupled Boltzmann equations involving two DM fermions as well as the late decaying singlet scalar. We have also shown how XENON1T data can be fitted by the boosted SIDM in this scenario.
		\begin{figure}[h!]
			\centering
			\includegraphics[scale=0.5]{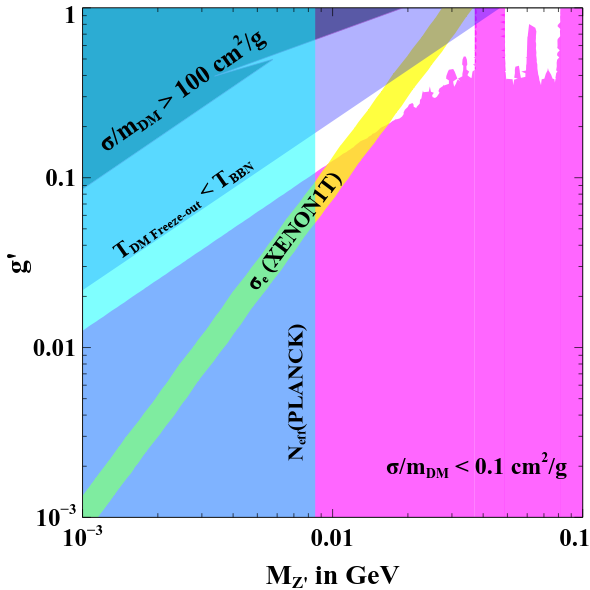}
			\caption{{Summary plot showing the parameter space in $g'-M_{Z'}$ plane for DM mass $m_{\rm DM}=0.1$ GeV.}}
			\label{summary1}
		\end{figure}
		
		\begin{figure}[h!]
			\centering
			\includegraphics[scale=0.5]{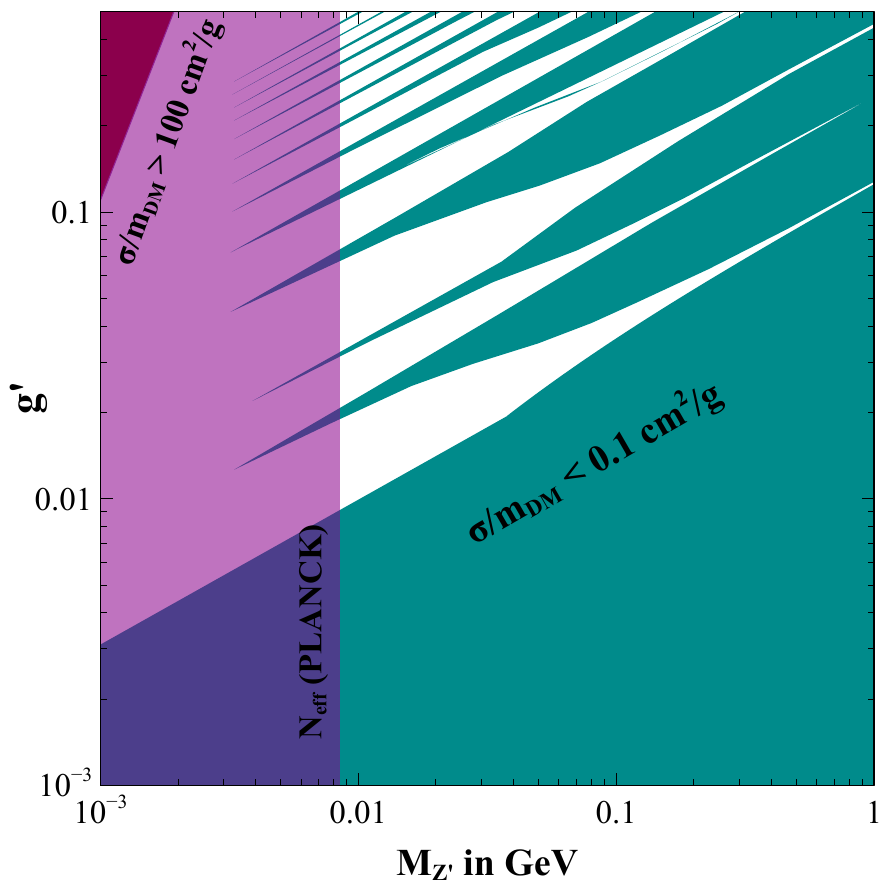}
			\caption{{Summary plot showing the parameter space in $g'-M_{Z'}$ plane for DM mass $m_{\rm DM}=100$ GeV.}}
			\label{summary2}
		\end{figure}
		
		In Fig.~\ref{summary1}, we summarize the final parameter space in $g'-M_{Z'}$ plane considering DM mass to be 0.1 GeV and $\epsilon=4.2\times10^{-4}$. Note that the chosen value of the kinetic mixing parameter is consistent with the latest constraints from  beam-dump and fixed target experiments, $e^{+}e^{-}$ colliders, lepton precision experiments as well as laboratory neutrino experiments which constrain such secluded $U(1)_{D}$ gauge boson~\cite{Bauer:2018onh}. The upper left and lower right regions are disfavored as they give rise to too large and too small DM self-interactions respectively, leaving a band in between. From this band also, more than half of the region is disfavored from the criteria of DM freeze-out happening before the BBN epochs. Although, by DM freeze-out, we mean DM freezing out within dark sector only where DM annihilates primarily into each other or light vector boson $Z'$; eventually, $Z'$ will decay into SM particles via kinetic mixing as it can not decay into DM kinematically. Therefore, as a conservative bound, we impose the criteria of DM freeze-out temperature to be more than BBN temperature. Very light $Z'$ is ruled out from cosmological constraints on effective relativistic degrees of freedom \cite{Planck:2018vyg, Kamada:2018zxi, Ibe:2019gpv, Escudero:2019gzq}. This arises due to the late decay of such light gauge bosons into SM leptons, after standard neutrino decoupling temperatures, thereby enhancing $N_{\rm eff}$. The corresponding disfavored region is shaded in cyan colour. Several constraints on such secluded gauge bosons arise from astrophysical observations. The constraint from white dwarf cooling is measured by observing variations of the white dwarf luminosity function and this arises because the plasmon inside the star can decay to neutrinos through this gauge boson,  leading to an enhanced cooling efficiency. However in case of a secluded hidden gauge boson, this contribution is strongly suppressed, because its coupling to neutrinos only arises through mixing with the $Z$ boson and hence does not constrain the parameter space shown in Fig~\ref{summary1} which is discussed in~\cite{Bauer:2018onh,Dreiner:2013tja}. Our parameter space is also safe with respect to the constraints from Supernova 1987A which arises because if such dark bosons are produced in sufficient quantity, they reduce the amount of energy emitted in the form of neutrinos, in conflict with observations~\cite{Chang:2016ntp}. Thus, only the thin white colored region on the upper right half of the plane remains allowed from these criteria. 
		The yellow coloured band denotes the required $\chi_2-e$ scattering cross-section to give rise to the XENON1T fit.  Clearly, only a tiny region remains allowed from all these criteria, which can be further scrutinized at near-future experiments. Note that the parameter scan done in Fig.~\ref{dd_con} for direct detection of boosted DM at DM-nucleon scattering experiments is consistent with this tiny region in Fig.~\ref{summary1}. It is noteworthy that, we are not incorporating DM relic constraints in this plane as those can be satisfied independently by appropriate tuning of singlet scalar couplings.It should be noted that we have chosen light sub-GeV DM in order to get the desired boosted DM flux as well as DM-electron scattering without conflicting other existing bounds. This has led to a very tiny allowed region of parameter space. To make this point clear, we also show another summary plot in Fig.~\ref{summary2} by considering DM mass to be 100 GeV. Clearly we have more allowed region of parameter space, although XENON1T fit is not possible in such a scenario.
		While we have confined ourselves to the discussion of DM aspects only in this work, such dark $U(1)_D$ gauge symmetry can also have consequences for the origin of light neutrino mass \cite{Borah:2020smw}, flavor anomalies \cite{Tsai:2019buq}, as well as cosmological phase transitions and gravitational waves \cite{Borah:2021ocu}. We leave such interesting aspects of $U(1)_D$ gauge symmetry to future studies.
		
		\acknowledgements
		DB acknowledges the support from Early Career Research Award from the Science and Engineering Research Board (SERB), Department of Science and Technology (DST), Government of India (reference number: ECR/2017/001873). MD acknowledges DST, Government of India for providing the financial assistance for the research under the grant 
		DST/INSPIRE/03/ 2017/000032. SM would like to thank K. Agashe for useful discussions.

		\appendix

		\section{Relevant cross section and decay widths}	
		\label{appen2}
		{\begin{eqnarray}
				\Gamma({\Phi_{1}\rightarrow \chi_1 \chi_1}) &=& \frac{y^2_1}{8\pi}m_{\Phi_1}\Big(1-4\frac{m^2_{\chi_1}}{m^2_{\Phi_1}} \Big)^{3/2}\\
				\sigma({\chi_1 \chi_1 \rightarrow \chi_2 \chi_2})&=&\frac{y_1'^2 y'^2_2}{32 \pi s }\frac{(s-4m^2_{\chi_2})^{3/2}(s-4m^2_{\chi_1})^{1/2}}{(s-m^2_{\Phi_2})^2}\\
				\sigma({\rm \chi \;\chi} \rightarrow Z' Z') &=& \frac{g'^4}{192 \pi s (s-4m^2_{\chi})} \times \Bigg[\frac{24s(4m^4_{\chi}+2M^4_{Z'}+sm^2_{\chi})A}{M^4_{Z'}+m^2_{\chi}s-4M^2_{Z'}m^2_{\chi}}\nonumber\\ & -&\frac{24 (8m^2_{\chi}-4M^2_{Z'}-s^2-(s-2M^2_{Z'})4m^2_{\chi})}{s-2M^2_{Z'}} {\rm Log}\Big[\frac{2M^2_{Z'}+s(A-1)}{2M^2_{Z'}-s(A+1)}\Big]\Bigg]\nonumber \\
			\end{eqnarray}
			where $A=\sqrt{\frac{(s-4M^2_{Z'})(s-4m^2_{\chi})}{s^2}}$}
		\begin{eqnarray}
			\sigma( e^{+} e^{-} \rightarrow {\rm \chi \;\chi})&=& \frac{g^2 g'^2 \epsilon^2(s+2m^2_{\chi})(s-m^2_e-4(s+2m^2_e)\sin^2\theta_{W})}{96 \pi \cos^2\theta_{W}(s-4m^2_e)(s-m^2_{Z'})^2}\sqrt{\frac{(s-4m^2_e)(s-4m^2_{\chi})}{s^2}} \nonumber \\
		\end{eqnarray}	
		
		Thermal averaged cross-section for annihilation of any particle $A$ to $B$  is given by: \cite{Gondolo:1990dk}
		\begin{equation}
			\langle\sigma v \rangle_{AA \rightarrow BB} = \frac{x}{2\big[K^2_1(x)+K^2_2(x)\big]}\times \int^{\infty}_{2}  dz \sigma_{(AA\rightarrow BB)}(z^2 - 4)z^2 K_1(zx)
			\label{appeneq1}
		\end{equation}
		where $z=\sqrt{s}/m_A$ and $x=m_A/T$.
		
		Thermal averaged decay width of $\Phi_1$ decaying to $\chi_1$ is given by:
		\begin{equation}
			\langle \Gamma(\Phi_1 \rightarrow \chi_1 \chi_1) \rangle = \Gamma(\Phi_1 \rightarrow \chi_1 \chi_1)  \Bigg(\frac{K_1(x)}{K_2(x)}\Bigg)
			\label{appeneq2}
		\end{equation}
		
		In Eq. \eqref{appeneq1} and \eqref{appeneq2}, $K_1$ and $K_2$ are the modified Bessel functions of 1st and 2nd kind respectively.

		\section{Evolution of DM Number Density over Galactic Time scale}	
		\label{mw_dm_evolution}	
		The boosted DM flux is inversely proportional to the mass square of the heavier DM component $\chi_1$ and is directly proportional to the annihilation cross-section of $\chi_1$ to $\chi_2$. For a chosen DM mass $m_{\chi_1}=0.1 $GeV, the correct boosted DM flux that can explain the excess electronic recoil events at XENON1T can be obtained if the annihilation cross-section $\langle\sigma_{ \chi_1\chi_1\rightarrow\chi_2\chi_2}v\rangle$ is $\mathcal{O}$($10^{-31}$cm$^2$) which is 5 orders larger than the typical WIMP annihilation cross-section. Hence it is imperative to check how does this annihilation of $\chi_1$ to $\chi_2$ affects their abundance in the Milky Way. This can be made certain by solving the following evolution equation for the DM number density:
		\begin{eqnarray}
			\frac{dn_{\chi_1}}{dt}&=&-\Gamma(\chi_1\chi_1\rightarrow\chi_2\chi_2)~ n_{\chi_1}
		\end{eqnarray}
		where $\Gamma(\chi_1\chi_1\rightarrow\chi_2\chi_2)$ is the interaction rate given by $n_{\chi_1}\langle\sigma_{ \chi_1\chi_1\rightarrow\chi_2\chi_2}v\rangle$.
		The solution of this equation gives:
		\begin{equation}
			n^{today}_{\chi_1}=\Bigg[\frac{1}{n^{init.}_{\chi_1}}-\langle\sigma_{ \chi_1\chi_1\rightarrow\chi_2\chi_2}v\rangle ~t_{\rm Milky Way}\Bigg]^{-1}
		\end{equation}
		where $t_{\rm MilkyWay} = 13.61$ Billion years($= 4.3\times10^{17}$ s.$=1.29\times 10^{28}$ cm.), is the age of the Milky Way galaxy and $n^{init.}_{\chi_1}$ is the number density of DM at the beginning of the formation of the Milky Way galaxy {\it i.e.}  $n^{init.}_{\chi_1}= \rho^{\rm DM}_{local}/2m_{\chi_1}$ (as $\Omega_{\chi_1}=\Omega_{\chi_2}=\Omega_{\rm DM}/2$). Our calculation shows that if $\langle\sigma_{ \chi_1\chi_1\rightarrow\chi_2\chi_2}v\rangle$ is smaller than $\mathcal{O}$($10^{-30}$)cm$^2$, then annihilation of $\chi_1$ to $\chi_2$ will neither change their abundance in Milky Way nor will the resulting flux affect the mass of the Milky Way by evaporation of DM.   
		


\end{document}